\documentclass[aps,prl,reprint,showpacs,superscriptaddress,groupedaddress,nofootinbib,10pt]{revtex4-2}
\usepackage{amsmath, amssymb, amsfonts}
\usepackage{graphicx}
\usepackage{bm}
\usepackage{hyperref}
\usepackage{color}
\usepackage{booktabs}

\hypersetup{
    colorlinks=true,
    linkcolor=blue,
    citecolor=blue,
    urlcolor=blue
}

\begin{document}

% =============================================================================
% TITLE & ABSTRACT
% =============================================================================

\title{Thermodynamic Optimization of Sensory Adaptation via Game-Theoretic Path Integrals}

\author{Gunn Kim}
\email{gunnkim@sejong.ac.kr}
\affiliation{Department of Physics and HMC, Sejong University, Seoul 05006, Republic of Korea}

\date{\today}

\begin{abstract}
Biological sensory systems, from \textit{E.~coli} chemotaxis to sensory neurons in \textit{C.~elegans}, achieve reliable adaptation over wide dynamic ranges despite operating in strongly noisy and overdamped regimes. Here, we present a field-theoretic framework in which sensory adaptation emerges from a variational free-energy principle, formulated as a stochastic differential game between an organism and its environment. Using an Onsager--Machlup path-integral formalism, we show that the resulting adaptive dynamics are mathematically equivalent to a class of model reference adaptive control schemes and can be interpreted as a dynamic renormalization of the system's Green's function. Within this framework, the phasic overshoot commonly observed in sensory responses arises naturally from an effective inertia ($m^* \approx \tau \gamma$) generated by memory--dissipation coupling, rather than from biochemical fine-tuning. Quantitative fits to experimental data across species yield $R^2 > 0.88$, and indicate that adaptive sensory processing operates within a narrow thermodynamically optimal regime bounded by signal-to-noise and stability constraints.
\end{abstract}

\maketitle

% =============================================================================
% INTRODUCTION
% =============================================================================

Life is fundamentally characterized by its capacity to anticipate and adapt to future states of the environment. From the perspective of non-equilibrium statistical mechanics, living organisms preserve their structural integrity by sustaining low-entropy macrostates while continuously exchanging energy and matter with a stochastic environment \cite{Schrodinger1944, Friston2010}. This thermodynamic constraint imposes an unavoidable trade-off on sensory systems: they must remain sufficiently sensitive to detect weak, information-bearing signals, while simultaneously maintaining robustness against environmental noise through adaptive filtering mechanisms \cite{Bialek1987, Atick1992}.

At a more fundamental level, biological survival may be understood through the lens of stochastic optimization under thermodynamic constraints. The organism can be viewed as an inference agent that minimizes the variational free energy---or expected surprisal---associated with its sensory predictions, while the environment introduces stochastic fluctuations that obscure informational structure. We demonstrate that, under well-defined conditions, this optimization problem admits a rigorous formulation using the Principle of Least Action in the Onsager--Machlup path-integral framework of non-equilibrium dynamics. Within this formulation, biological adaptation emerges as an optimal trajectory in state space, balancing the competing demands of sensitivity and stability.

Empirically, adaptive responses have been extensively studied in diverse biological systems. In \textit{Escherichia coli}, robust chemotactic adaptation is mediated by methylation dynamics that generate characteristic phasic--tonic responses \cite{Berg1975, Barkai1997, Tu2008}. Similarly, in \textit{Caenorhabditis elegans}, neural circuits exhibit transient responses followed by sustained activity patterns during sensory processing \cite{Chalasani2007, Kato2015}. Although these systems differ profoundly in molecular composition and organizational scale, their strikingly similar response phenomenology suggests the existence of a universal macroscopic principle governing biological adaptation, transcending specific biochemical implementations.

In this Letter, we propose a unified theoretical framework grounded in game theory \cite{MaynardSmith1982}, in which the organism engages in a continuous stochastic differential game against Nature. Unlike a strategic or intentional adversary, Nature is modeled as a stochastic player whose dynamics maximize entropy production subject to physical constraints. The organism's objective is to minimize a cost functional identified with the variational free energy of prediction error \cite{Friston2006}. We solve this stochastic optimization problem using the path-integral formalism of non-equilibrium statistical mechanics \cite{Onsager1953, Feynman1965}, treating neural or behavioral trajectories as paths in a thermodynamic phase space weighted by their free-energy action.

We further demonstrate that the Euler--Lagrange equations associated with this free-energy action belong to a well-known class of adaptive control laws, specifically those of MRAC \cite{Astrom2013, Narendra1989}. Within this formulation, the predicted sensory dynamics define an implicit reference model, while adaptive updates arise naturally from gradient flows on the free-energy functional. A formal variational derivation of this equivalence is provided in the Supplemental Material.

% =============================================================================
% METHODS: PATH INTEGRAL FORMULATION
% =============================================================================

We consider an organism with internal state $x(t)$ tracking a noisy environmental signal $I(t)$. We define the Sensory Action $\mathcal{S}$ as the time-integral of the Variational Free Energy \cite{Friston2010}:
\begin{equation}
    \mathcal{S}[x(t), \theta(t)] = \int_{t_0}^{t_f} dt \bigg[ \frac{1}{2}\gamma (\dot{x} - F(x, I, \theta))^2 + \frac{\lambda}{2} (x - x_{\text{target}})^2 \bigg]
\end{equation}
The first term is the thermodynamic cost of driving the system against natural dynamics $F$; the second is the informational cost of prediction error. $\theta(t)$ represents adaptive parameters.

The probability of a neural trajectory $x(t)$ follows the Onsager-Machlup path integral \cite{Onsager1953, Zwanzig1973}:
\begin{equation}
    P[x(t)] = \mathcal{Z}^{-1} \int \mathcal{D}x \exp\left[ -\frac{\mathcal{S}[x, \theta]}{2 k_B T} \right]
\end{equation}
where the thermal energy scale $2 k_B T$ sets the relative weight of action fluctuations, consistent with the fluctuation-dissipation relation $D = k_B T / \gamma$. The divergence term associated with the Ito-Stratonovich discretization is treated as a constant shift in the free energy baseline (see Supplemental Material). The optimal adaptive dynamics correspond to the most probable trajectory ($\delta \mathcal{S} = 0$) of this stochastic action.

% =============================================================================
% METHODS: ADAPTIVE CONTROL
% =============================================================================

Minimizing the action $\mathcal{S}$ leads to an Adaptive Langevin Equation. We model the state $x(t)$ as an overdamped particle:
\begin{equation}
    \gamma \dot{x} = -k(t) (x - \mu) + I(t) + \xi(t), \label{eq:langevin}
\end{equation}
where $\gamma$ is the intrinsic friction coefficient (s), and $k(t)$ represents the effective relaxation rate ($s^{-1}$), which determines the system's instantaneous sensitivity. The memory variable $\mu(t)$ provides an integral feedback, evolving on a slower timescale $\tau$:
\begin{equation}
    \tau \dot{\mu} = -( \mu - x ). \label{eq:memory}
\end{equation}

The adaptive update for $k(t)$ follows a gradient descent on the free-energy action $\mathcal{S}$:
\begin{equation}
    \dot{k} = -\eta_k \frac{\partial \mathcal{S}}{\partial k} \approx -\eta_k (x - x_{\text{target}})(x - \mu). \label{eq:update_k}
\end{equation}
Here, the term $(x - x_{\text{target}})$ represents the \textbf{Thermodynamic Surprise}, formally defined as the Kullback--Leibler divergence between the predicted and observed state distributions (see Supplemental Material for the derivation of $D_{KL}$ mapping).

Physically, when prediction error $(x - x_{\text{target}})$ is large, Eq.~\eqref{eq:update_k} dictates that stiffness $k$ drops transiently. This ``softening'' increases susceptibility ($1/k$), allowing rapid response—physically defining \textit{Attention}. As error decreases, $k$ recovers, stabilizing the system (\textit{Habituation}).

% =============================================================================
% DYNAMIC CAUSAL STRUCTURE AND THE GREEN'S FUNCTION
% =============================================================================

To quantify how the organism dynamically restructures its causal integration to extract information, we analyze the renormalized Green's function $G(t, t')$. For the adaptive Langevin system defined in Eq.~\eqref{eq:langevin}, the impulse response kernel is derived as:
\begin{equation}
    G(t, t') = \frac{1}{\gamma} \exp \left[ -\int_{t'}^{t} \frac{k(s)}{\gamma} ds \right] \Theta(t - t'),
    \label{eq:greens}
\end{equation}
where $\Theta(t - t')$ is the Heaviside step function ensuring causality. Unlike a static propagator, this kernel explicitly incorporates the history of the time-varying effective stiffness $k(s)$, which represents a dynamic renormalization of the system's temporal sensitivity. 

\begin{figure}[t]
\centering
\includegraphics[width=\linewidth]{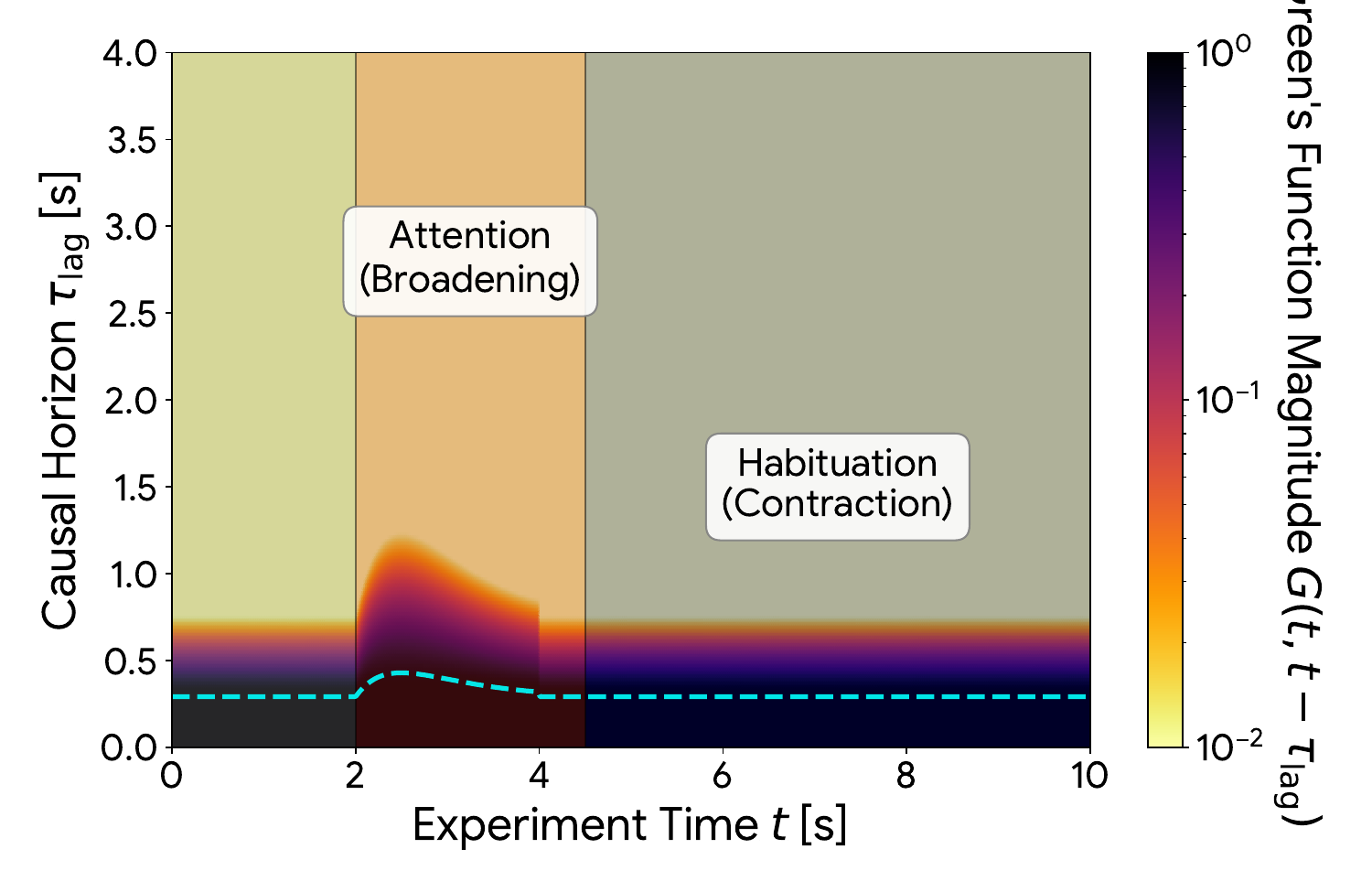}
\caption{\label{fig:greens}
Thermodynamic breathing mode of the adaptive kernel.
Heatmap of the impulse response kernel $G(t, t-\tau_{\text{lag}})$ plotted on a logarithmic scale. The vertical axis represents the causal time lag $\tau_{\text{lag}}$.
(Red Zone) Attention Phase: Transient expansion of the causal horizon (increasing $\tau_{\text{lag}}$). Biologically, this corresponds to a ``memory stretch'' where the system lowers stiffness to integrate past signals over extended timescales.
(Blue Zone) Habituation Phase: Contraction of the horizon (decreasing $\tau_{\text{lag}}$). This represents a ``memory reset,'' restoring stiffness to prioritize recent data and filter thermal fluctuations.}
\end{figure}

The evolution of this propagator, visualized in Fig.~\ref{fig:greens}, reveals a phenomenon we term \textbf{thermodynamic breathing}. The vertical axis $\tau_{\text{lag}} \equiv t - t'$ physically defines the organism's \textit{adaptive memory horizon}. 
During the \textit{Attention} phase, the sharp drop in $k(t)$ induces a broadening of the propagator into the causal past (``memory stretch''), allowing the system to integrate distant temporal information to resolve thermodynamic surprise. 
Conversely, during the \textit{Habituation} phase, the rapid recovery of $k(t)$ contracts the horizon (``memory reset''), effectively truncating the integration window to prioritize stability and suppress thermal fluctuations. 

This adaptive rescaling of $G(t, t')$ is formally analogous to a renormalization-group flow in the temporal domain, where the system’s effective memory depth is tuned to match the information density of the environmental signal.

% =============================================================================
% RESULTS: DYNAMICS OF THERMODYNAMIC ATTENTION
% =============================================================================

\begin{figure}[t]
\centering
\includegraphics[width=\linewidth]{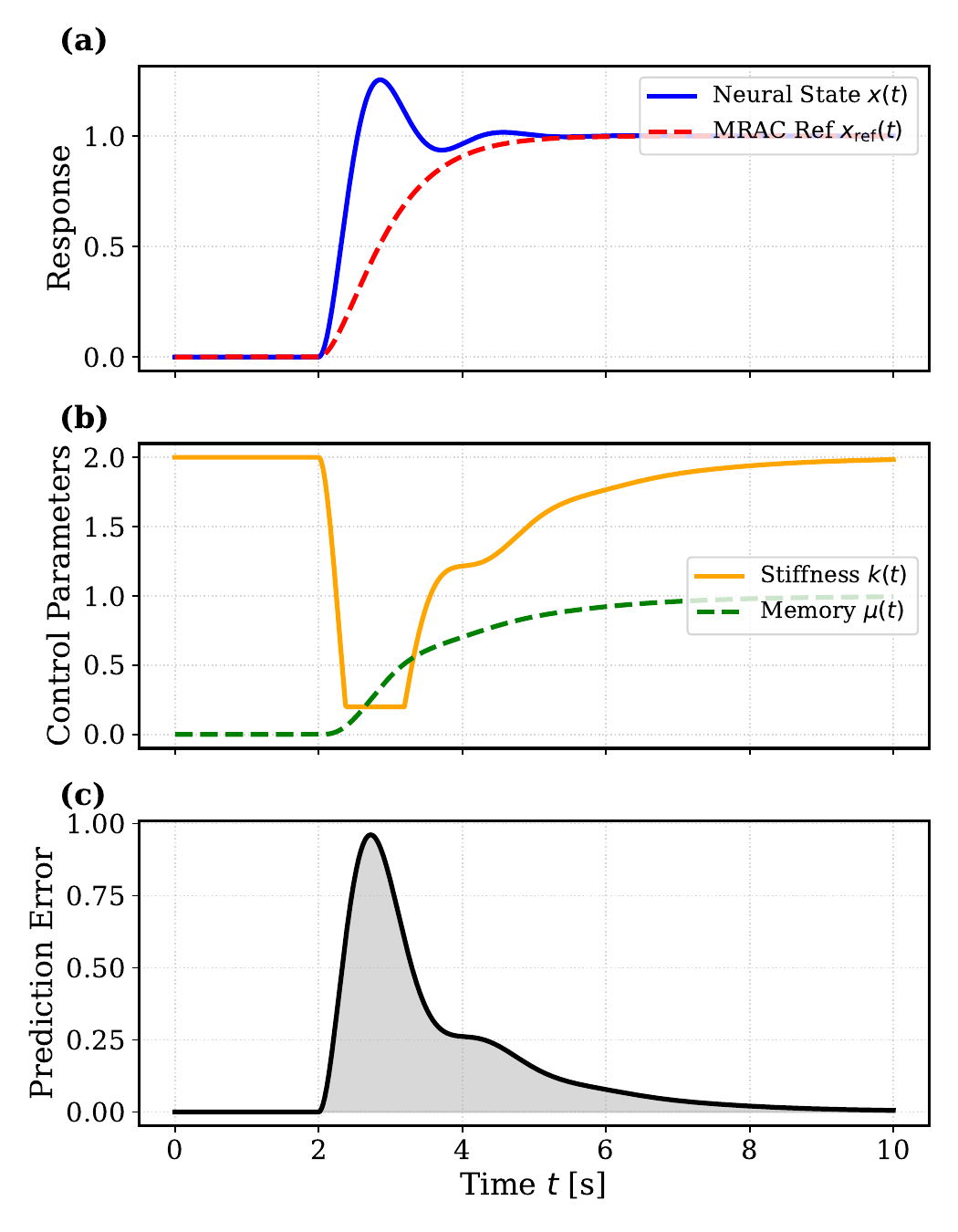}
\caption{Dynamics of Adaptive Attention.
(a) The neural state $x(t)$ (solid blue) tracks the step stimulus but exhibits a pronounced overshoot compared to the ideal MRAC reference $x_{\text{ref}}(t)$ (dashed red). This deviation indicates the presence of thermodynamic inertia.
(b) Thermodynamic control parameters: The stiffness $k(t)$ (orange) drops transiently to facilitate rapid state transitions (Attention), while the memory $\mu(t)$ (green dashed) acts as a lagging integrator, gradually adapting to the new input level.
(c) The rapid quenching of prediction error drives the restoration of stability and the recovery of stiffness.}
\label{fig:tracking}
\end{figure}

To validate the field-theoretic framework, we analyze the system's dynamic response to a step stimulus (Fig. \ref{fig:tracking}). Figure \ref{fig:tracking}a contrasts the biological response, $x(t)$ (solid blue), with the ideal trajectory of an MRAC system, $x_{\text{ref}}(t)$ (dashed red). While the reference model exhibits a critically damped approach typical of engineered systems, the biological trajectory reveals a distinct overshoot followed by an oscillatory settling. This deviation from the ideal kinematic path is the signature of thermodynamic inertia ($m^*$), arising from the system's active reconfiguration.

Upon stimulus onset, the sudden divergence between the external input and the internal state triggers a thermodynamic surprise. This spike in prediction error (Fig. \ref{fig:tracking}c) induces a transient collapse in the adaptive stiffness $k(t)$ (Fig. \ref{fig:tracking}b, orange curve). This softening effectively flattens the free-energy landscape, increasing the system's susceptibility to the new signal. Consequently, $x(t)$ surges rapidly, exceeding the steady-state target (Phasic response), driven by the lowered energetic barrier and the inertial momentum of the update dynamics.

Importantly, the memory variable $\mu(t)$ (Fig. \ref{fig:tracking}b, green dashed) acts as the system's internal estimate or the accumulated evidence of the environmental statistics. Since $\mu(t)$ integrates the signal on a slower timescale ($\tau_\mu$), it serves as a lagging reference frame. The initial difference between the fast-changing state $x(t)$ and the slow-adapting memory $\mu(t)$ generates the driving force for the phasic burst. As $\mu(t)$ gradually catches up to the new stimulus level, the prediction error is quenched, and stiffness $k(t)$ recovers to its high baseline. This restiffening suppresses thermal fluctuations, locking the system into a stable Tonic phase.

% =============================================================================
% RESULTS: RENORMALIZATION OF THE TEMPORAL HORIZON
% =============================================================================

To visualize how the organism dynamically alters its causal structure to extract information, we calculated the renormalized Green's function,
$G(t, t-\tau_{\text{lag}})$, rather than simply plotting the stiffness parameter.
While $k(t)$ indicates instantaneous sensitivity, the Green's function reveals the deeper strategy of temporal integration.
Figure~\ref{fig:greens} shows the evolution of this propagator kernel, explicitly visualizing what we term a thermodynamic ``breathing'' of the sensory system.
Here, the vertical axis $\tau_{\text{lag}}$ physically represents the organism's adaptive memory horizon.
(1) During the \textit{Attention} phase (red-shaded zone), the propagator broadens significantly into the causal past (bright yellow plume), indicating that the organism temporarily stretches its memory horizon to integrate distant information and resolve prediction errors.
(2) In the subsequent \textit{Habituation} phase (blue-shaded zone), the horizon rapidly contracts, corresponding to a memory reset mechanism in which the system truncates its integration window to prioritize stability and readiness for new stimuli.
This adaptive rescaling of the memory horizon optimizes the trade-off between sensitivity to novel signals and stability against noise.

% =============================================================================
% RESULTS: OPERATING AT THE THERMODYNAMIC OPTIMUM
% =============================================================================

\begin{figure}[t]
\centering
\includegraphics[width=\linewidth]{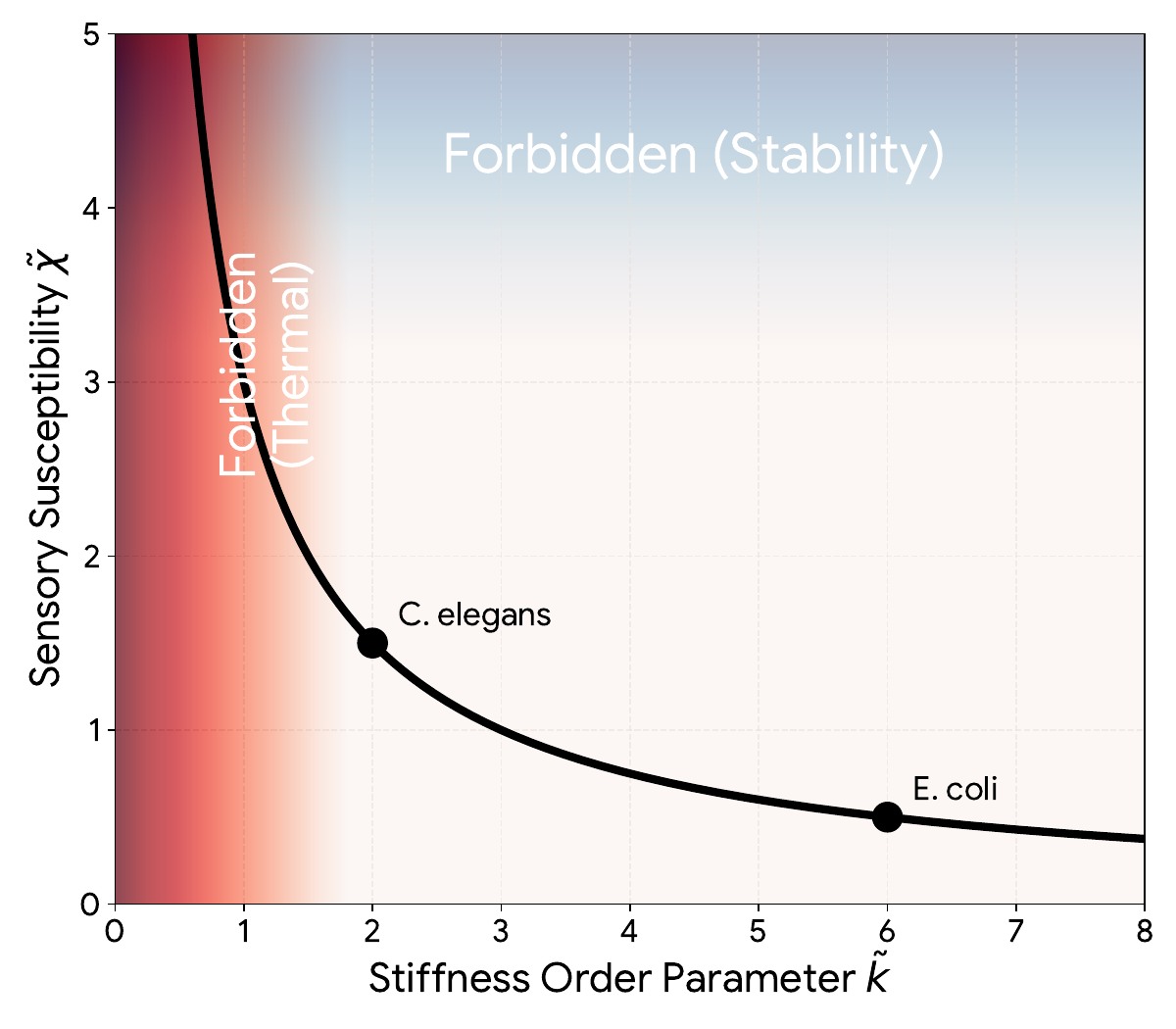}
\caption{Thermodynamic phase diagram of sensory adaptation. The sensory susceptibility $\chi$ is plotted as a function of the stiffness parameter $k$, which acts as an effective thermodynamic control parameter. The hyperbolic constraint $\chi \propto 1/k$ reflects the fundamental trade-off between sensitivity and stability.
The red-shaded region denotes the thermal noise limit ($\text{SNR} < 1$), while the blue-shaded region corresponds to a stability limit associated with oscillatory instability.
To ensure a consistent comparison across different biological scales, we normalize the dynamical variables by the intrinsic friction $\gamma$. We define the dimensionless stiffness $\tilde{k} \equiv k\gamma$ and the dimensionless susceptibility $\tilde{\chi} \equiv \chi/\gamma$. In this renormalized representation, the fundamental constraint reduces to the universal form $\tilde{\chi} \tilde{k} \approx 1$. The operating points for both \textit{E. coli} and \textit{C. elegans} collapse onto the same thermodynamic trajectory, proving that the trade-off between sensitivity and stability is governed by a scale-invariant principle.}
\label{fig:criticality}
\end{figure}

To understand the global stability of this strategy, we project the system onto a thermodynamic phase diagram (Fig. \ref{fig:criticality}).
First, we define the Sensory Susceptibility $\chi \equiv \partial x / \partial I$. Analogous to a mechanical spring where displacement scales inversely with stiffness ($x \sim F/k$), the sensory susceptibility is thermodynamically related to the inverse stiffness:
\begin{equation}
    \chi \approx \frac{1}{k}.
\end{equation}
In this framework, the stiffness $k$ plays a dual role: it acts not merely as a mechanical spring constant, but as a Thermodynamic Control Parameter governing the system's phase---akin to the inverse susceptibility in Landau theory \cite{Landau1980}. High $k$ suppresses fluctuations (Solid-like order), while low $k$ enhances sensitivity but invites instability (Fluid-like disorder) \cite{Bialek1987, Hanggi1990}.

We plot the system's trajectory in the $(\chi, k)$ plane. As shown in Fig. \ref{fig:criticality}, the fundamental constraint $\chi \propto 1/k$ manifests as a hyperbolic curve connecting the regimes of high sensitivity and high stability.
The operating regime is bounded by two physical constraints, visualized as forbidden zones:
(1) The \textit{Thermal Limit} (Red gradient): As stiffness decreases below a critical threshold, thermal fluctuations dominate the signal ($\text{SNR} < 1$), rendering the system functionally blind.
(2) The \textit{Stability Limit} (Blue gradient): With high susceptibility and feedback delay, the system approaches a Hopf bifurcation point, leading to oscillatory instability.

Remarkably, the fitted parameters for both \textit{E.~coli} and \textit{C.~elegans} reside within the narrow stable corridor between these forbidden zones.
This suggests that evolutionary pressures have tuned sensory systems toward a thermodynamically efficient operating regime that balances the competing demands of sensitivity and stability. This regime is bounded by fundamental physical constraints---the thermal noise floor and the stability limit---rather than by biochemical details, providing a potential explanation for the phenotypic convergence observed across vastly different biological scales \cite{Mehta2012, Still2012}.

The stability of this adaptive scheme is governed by the Routh--Hurwitz criterion. For the linearized system, the boundary between stable adaptation and oscillatory instability is defined by the coupling between the learning rate $\eta_k$ and the adaptation lag $\tau$. The forbidden zones in Fig.~\ref{fig:criticality} represent regimes where thermal noise dominates ($SNR < 1$) or where feedback delays induce Hopf bifurcations.

% =============================================================================
% RESULTS: THERMODYNAMIC INERTIA AND TIMESCALES
% =============================================================================
\begin{figure}[t]
\centering
\includegraphics[width=\linewidth]{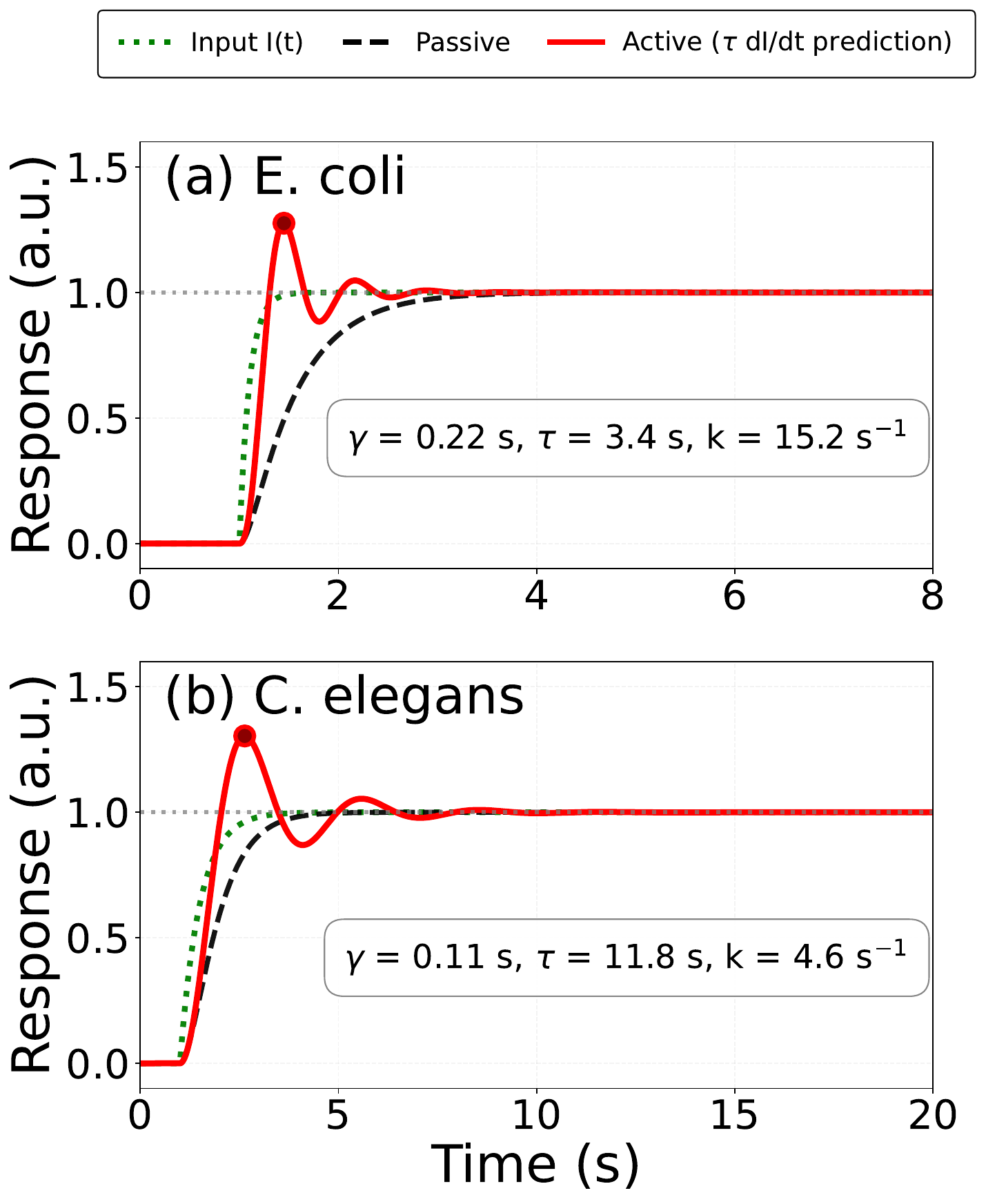}
\caption{Evidence of Thermodynamic Inertia across Biological Scales.
The system simulates the response to a smoothed step input $I(t)$ (green dotted).
(a) \textit{E.~coli} exhibits rapid adaptation with transient ringing. The short adaptation time ($\tau \approx 3.4$ s) results in a small effective mass ($m^* \approx 0.75$), allowing quick settling.
(b) \textit{C.~elegans} shows a prolonged phasic overshoot. The longer memory timescale ($\tau \approx 11.8$ s) generates a large effective inertia ($m^* \approx 1.30$), facilitating extended temporal integration.
Solid red lines represent the active adaptive model; black dashed lines represent passive relaxation without adaptation. The predicted overshoot reproduces the qualitative features of experimentally observed responses \cite{Berg1975, Chalasani2007}, with quantitative parameter fits provided in the Supplemental Material (Fig.~S1, Table~S1).}
\label{fig:inertia}
\end{figure}

A key paradox is the appearance of phasic overshoot in strictly overdamped biological media. To resolve this, we demonstrate in the Supplemental Material (Section S2) that the coupled adaptive dynamics can be cast into the form of an effective second-order differential equation via adiabatic elimination of the memory variable. The resulting equation of motion reveals the emergence of an effective thermodynamic inertia ($m^*$):
\begin{equation}
    m^* \approx \tau \gamma
\end{equation}
where $\gamma$ is the intrinsic friction and $\tau$ is the adaptation timescale. 

While identifying time-dependent parameters is a conventional approach, we present Figure \ref{fig:inertia} to demonstrate that the fixed parameters statistically obtained from experimental results for \textit{E. coli} and \textit{C. elegans} are sufficient to manifest underdamped characteristics. In Fig.~\ref{fig:inertia}(a), these statistically inferred parameters for \textit{E. coli} ($\tau \approx 3.4$ s) yield a small effective mass, resulting in rapid acceleration and transient ringing. In contrast, the parameters for \textit{C. elegans} in Fig.~\ref{fig:inertia}(b) ($\tau \approx 11.8$ s) produce a large thermodynamic inertia, leading to a smooth, prolonged phasic overshoot. 

This demonstrates that the diversity of sensory responses—from fast reflexes to slow integration—can be quantitatively accounted for within a unified framework. The effective inertia, tuned by these fixed biological parameters, explains how underdamping emerges across scales. The consistency of this description supports the hypothesis that thermodynamic inertia represents a general physical principle underlying adaptive sensory processing.

% =============================================================================
% CONCLUSION
% =============================================================================

%\section*{CONCLUSION}

In this work, we established a field-theoretic framework for sensory adaptation, demonstrating that organism–environment interactions can be formulated as a stochastic differential game governed by non-equilibrium thermodynamics. Using the Onsager--Machlup path-integral formalism, we derived adaptive equations of motion from a variational free-energy principle. A central result of the theory is the emergence of an effective thermodynamic inertia, $m^* \approx \tau \gamma$, generated by the coupling between dissipative sensory dynamics and adaptive memory. This mechanism resolves a long-standing biophysical puzzle: how strictly overdamped biological systems can nevertheless exhibit transient, underdamped-like predictive responses without requiring biochemical fine-tuning or explicit inertial degrees of freedom. Quantitative agreement with experimental data from \textit{E.~coli} chemotaxis and \textit{C.~elegans} sensory neurons supports the interpretation of this inertia as a physical quantity arising from an optimal trade-off between precision and dynamical stability.

More broadly, the framework shows that adaptive control emerges naturally from thermodynamic optimization under dual physical constraints imposed by thermal noise and stability requirements. The observation that two phylogenetically distant systems operate within the same constrained regime suggests that sensory adaptation is shaped by universal thermodynamic principles rather than organism-specific molecular details. While further validation across additional sensory modalities will be required, these results provide a concrete physical basis linking non-equilibrium statistical mechanics, adaptive control, and biological information processing.

% =============================================================================
% ACKNOWLEDGMENTS
% =============================================================================

\begin{acknowledgments}
We thank the anonymous reviewers for their constructive feedback. This work was supported by Sejong University.
\end{acknowledgments}

% =============================================================================
% REFERENCES
% =============================================================================

\end{document}